\documentclass{article}
\usepackage{ijcai20}

\usepackage{times}

\usepackage{soul}
\usepackage{url}
\usepackage[hidelinks]{hyperref}
\usepackage[utf8]{inputenc}
\usepackage[small]{caption}
\usepackage{graphicx}
\usepackage{amsmath}
\usepackage{booktabs}
\usepackage{algorithm}
\usepackage{algorithmic}

\usepackage{amssymb}
\usepackage{booktabs}
\usepackage{multirow}
\usepackage{subfigure}
\DeclareMathOperator{\sigmoid}{sigmoid}
\DeclareMathOperator{\FClayer}{FClayer}
\DeclareMathOperator{\BiLSTM}{BiLSTM}
\DeclareMathOperator{\Tr}{Tr}
\DeclareMathOperator{\WaveNet}{WaveNet2D}

\title{Robust Front-End for Multi-Channel ASR using Flow-Based Density Estimation}

\author{
Hyeongju Kim\footnote{Contact Author}\and
Hyeonseung Lee\and
Woo Hyun Kang\and
Hyung Yong Kim\And
Nam Soo Kim\\
\affiliations
Department of Electrical and Computer Engineering and INMC, Seoul National University, South Korea\\
\emails
\{hjkim, hslee, whkang, hykim\}@hi.snu.ac.kr,
nkim@snu.ac.kr
}

\begin{document}

\maketitle

\begin{abstract}
For multi-channel speech recognition, speech enhancement techniques such as denoising or dereverberation are conventionally applied as a front-end processor. Deep learning-based front-ends using such techniques require aligned clean and noisy speech pairs which are generally obtained via data simulation. Recently, several joint optimization techniques have been proposed to train the front-end without parallel data within an end-to-end automatic speech recognition (ASR) scheme. However, the ASR objective is sub-optimal and insufficient for fully training the front-end, which still leaves room for improvement. In this paper, we propose a novel approach which incorporates flow-based density estimation for the robust front-end using non-parallel clean and noisy speech. Experimental results on the CHiME-4 dataset show that the proposed method outperforms the conventional techniques where the front-end is trained only with ASR objective.
\end{abstract}

\section{Introduction}
Robust multi-channel speech recognition is a challenging task since the acoustic interferences such as background noise and reverberation degrade the quality of input speech. It is known that an automatic speech recognition (ASR) system, which is trained on clean speech, works poorly in noisy environments due to the mismatch in acoustic characteristics~\cite{gong1995speech}. For robust multi-channel ASR, recent studies usually employ a front-end component that leverages a denoising algorithm such as the Minimum Variance Distortionless Response (MVDR) or a dereverberation algorithm (e.g., the Weighted Prediction Error, WPE)~\cite{barker2017third,kinoshita2016summary}. Even though these denoising and dereverberation methods have brought substantial improvements for an ASR system~\cite{vincent2013second}, they are usually designed for enhancing speech in stationary environments.

For handling more realistic acoustic environments, speech enhancement techniques based on deep neural network (DNN) have been developed, which basically require time-aligned parallel clean and noisy speech data for training~\cite{erdogan2016improved}. These techniques usually train the model to optimize signal level criteria such as signal to noise ratio (SNR), independently of speech recognition accuracy. To make the speech enhancement algorithm a more efficient front-end for ASR, recent studies have proposed to optimize the speech enhancement model using the ASR objective within an end-to-end ASR scheme~\cite{heymann2019joint,ochiai2017multichannel,kinoshita2017neural}. This training method allows to use non-parallel clean and noisy data for training the front-end.

However, since the ASR objective only focuses on preserving the phonetic information of the input speech, it is insufficient for fully training the speech enhancement model and gauranteeing generalized performance improvement. Moreover, conventional approaches do not take into account the distribution of the target clean speech signal on which the original ASR system is trained. To overcome these limitations, we propose a novel method that applies flow-based density estimation to the robust front-end using non-parallel clean and noisy speech. In the proposed method, a flow-based density estimator is trained with clean speech and the front-end receives the additional generative loss from the density estimator. In other words, the front-end performs multi-task learning. The auxiliary objective induces the front-end to learn more regularized representations, which in turn improves the performance of the ASR module on the noisy CHiME-4 evaluation set.

Our main contributions are as follows:
\begin{itemize}
    \item We propose a novel approach that combines density estimation with  multi-channel ASR to exploit the probability distribution of speech signal for robust front-end.
	\item We present a new flow-based model \textit{MelFlow} for estimating the probability distribution of mel-spectrograms.
	\item We demonstrate that our multi-task learning strategy results in better performance on WERs over noisy speech compared to the scheme that depends only on the ASR objective.
\end{itemize}

\section{Related Work}
Denoising and dereverberation methods are originally designed to estimate clean signal and can be applied without any training~\cite{wolfel2009distant}. These traditional methods, however, require time-consuming iterative process and work well only in stationary environmental conditions. To handle more realistic acoustic environments, modern speech enhancement techniques usually employ DNNs to remove background noise and directly estimate the desired signal~\cite{han2015learning}. 

For multi-channel ASR, denoising and dereverberation techniques have been employed as front-ends and reported to produce some improvements in noisy speech recognition~\cite{wolfel2009distant}. However, the direct application of enhancement-based algorithms to ASR has some problems. One of the main problems is that the ASR accuracy is not taken into consideration when training the front-end, thus the resulting features may lack phonetic information. 
Another critical problem is that training the conventional enhancement modules require parallel dataset (i.e., pairs of aligned clean and noisy speech signals). 
To alleviate the problems, recent approaches optimize the front-end and ASR models jointly using the ASR objective~\cite{heymann2019joint,ochiai2017multichannel}.

\section{Baseline}
\subsection{Neural Beamforming Method}
A filter-and-sum beamforming method is a typical denoising technique for enhancing multi-channel signal. In the filter-and-sum beamforing, a speech image at the reference microphone is estimated by using a linear filter operating as follows:
\begin{equation}
\label{eq:eq1}
{y}_{t, f} =  \sum_{c=1}^{C}{h_{t,f,c}s_{t,f,c}},
\end{equation}
where $s_{t,f,c} \in \mathbb{C}$ is short-time Fourier transform (STFT) coefficient, $h_{t,f,c} \in \mathbb{C}$ is a beamforming filter coefficient and $y_{t,f} \in \mathbb{C}$ is an estimated speech image. Subscripts $t,f,c$ denote the $c$-th channel of a signal at a time-frequency bin $(t,f)$.
While conventional methods optimize $h_{t,f,c}$ based on a signal-level objective, recent studies train $h_{t,f,c}$ jointly within an ASR architecture~\cite{meng2017deep,ochiai2017multichannel}. This kind of data-driven approach is called the neural beamforming method and can be classified into two categories: (i) filter estimation approach and (ii) mask estimation approach. The filter estimation approach estimates the time-variant filter coefficients $\{h_{t,f,c}\}_{t=1,f=1,c=1}^{T,F,C}$ directly but suffers from unstable training due to high flexibility~\cite{meng2017deep}. On the other hand, the mask estimation approach optimizes time-invariant filter coefficients $\{h_{f,c}\}_{f=1,c=1}^{F,C}$ and has been reported to achieve improved performances in multi-channel speech recognition~\cite{ochiai2017multichannel}. Also, the mask estimation approach can be applied to any microphone configurations. Given the advantages of the latter, this paper focuses on the mask estimation approach.

\paragraph{Mask estimation approach.}
To get time-invariant coefficients $\{h_{f,c}\}_{f=1,c=1}^{F,C}$, we first calculate a speech mask $m^S_{t,f,c} \in [0,1]$. An input feature $\boldsymbol{\Tilde{s}}_{t,c} \in \mathbb{R}^F$ is the aggregation of the amplitudes of the $c$-th channel's time-frequency bin along the frequency axis at time $t$:
\begin{equation}
\label{eq:eq2}
\boldsymbol{\Tilde{s}}_{t,c}=\{\sqrt{\Re(s_{t,f,c})^2+\Im(s_{t,f,c})^2}\}^F_{f=1}.
\end{equation}
The speech mask $m^S_{t,f,c}$ is acquired from the input feature $\boldsymbol{\Tilde{s}}_{t,c}$ as follows:
\begin{equation}
\label{eq:eq3}
\{\boldsymbol{o}^S_{t,c}\}^T_{t=1} = \BiLSTM(\{\boldsymbol{\Tilde{s}}_{t,c}\}^T_{t=1};\theta_S),
\end{equation}
\begin{equation}
\label{eq:eq4}
\{m^S_{t,f,c}\}^F_{f=1} = \sigmoid(\FClayer(\boldsymbol{o}^S_{t,c};\phi_S)),
\end{equation}
where BiLSTM is a real-valued bidirectional LSTM network, $\boldsymbol{o}^S_{t,c} \in \mathbb{R}^{D_{out}}$ is the output of BiLSTM and FClayer is a fully connected network from $\mathbb{R}^{D_{out}} \mapsto \mathbb{R}^F$. A cross-channel power spectral density (PSD) matrix $\boldsymbol{\Phi}^S_f \in \mathbb{C}^{C \times C}$ for a speech signal can be obtained with a channel-averaged mask $m^S_{t,f}$ as follows:
\begin{equation}
\label{eq:eq5}
m^S_{t,f} = {\frac{1}{C}} \sum^C_{c=1}m^S_{t,f,c},
\end{equation}
\begin{equation}
\label{eq:eq6}
{\Phi}^S_f = {\frac{1}{\sum^T_{t=1}m^S_{t,f}}} \sum^T_{t=1}m^S_{t,f}\boldsymbol{s}_{t,f}\boldsymbol{s}_{t,f}^\dag,
\end{equation}
where $\boldsymbol{s}_{t,f} = \{s_{t,f,c}\}^C_{c=1} \in \mathbb{C}^C$ is the channelwise concatenated vector of the STFT coefficients and $\dag$ represents Hermitian transpose. Using the same architecture with different parameters $\theta_N$, $\phi_N$, another PSD matrix $\boldsymbol{\Phi}^N_f \in \mathbb{C}^{C \times C}$ for a noise signal is derived in the same way. Finally, the time-invariant linear filter coefficient $h_{f,c}$ is computed with the MVDR formulation as follows:
\begin{equation}
\label{eq:eq7}
\{h_{f,c}\}^C_{c=1}=\frac{{{\Phi}^N_f}^{-1}{\Phi}^S_f}{\Tr({{\Phi}^N_f}^{-1}{\Phi}^S_f)}\boldsymbol{r},
\end{equation}
where $\Tr (\cdot)$ is the trace operator and $\boldsymbol{r} \in \mathbb{R}^C$ is the one-hot vector indicating the index of a reference microphone. We can integrate another network to estimate the reference microphone in case the index of the reference is not specified and not available.

\subsection{ESPnet}
ESPnet~\cite{watanabe2018espnet} is an end-to-end ASR which is based on connecionist temporal classification (CTC) and attention mechanism. ESPnet has an attention-based encoder-decoder structure and shares encoder representations to optimize both CTC and attention-based cross entropy objectives jointly. This joint multi-task learning framework has been known to improve performance and achieve fast convergence~\cite{kim2017joint}. ESPnet also incorporates the neural beamforming method as a pre-processor and optimizes the front-end within the end-to-end ASR scheme. For decoding, the weighted average of attention-based and CTC scores is used to eliminate irregular alignments. We use ESPnet as a base ASR architecture and integrate a density estimator into the multi-channel ASR in the next section.

\section{Proposed Model}

\begin{figure*}[t]
	\centering
	\subfigure[]{
	\includegraphics[width=0.7\linewidth]{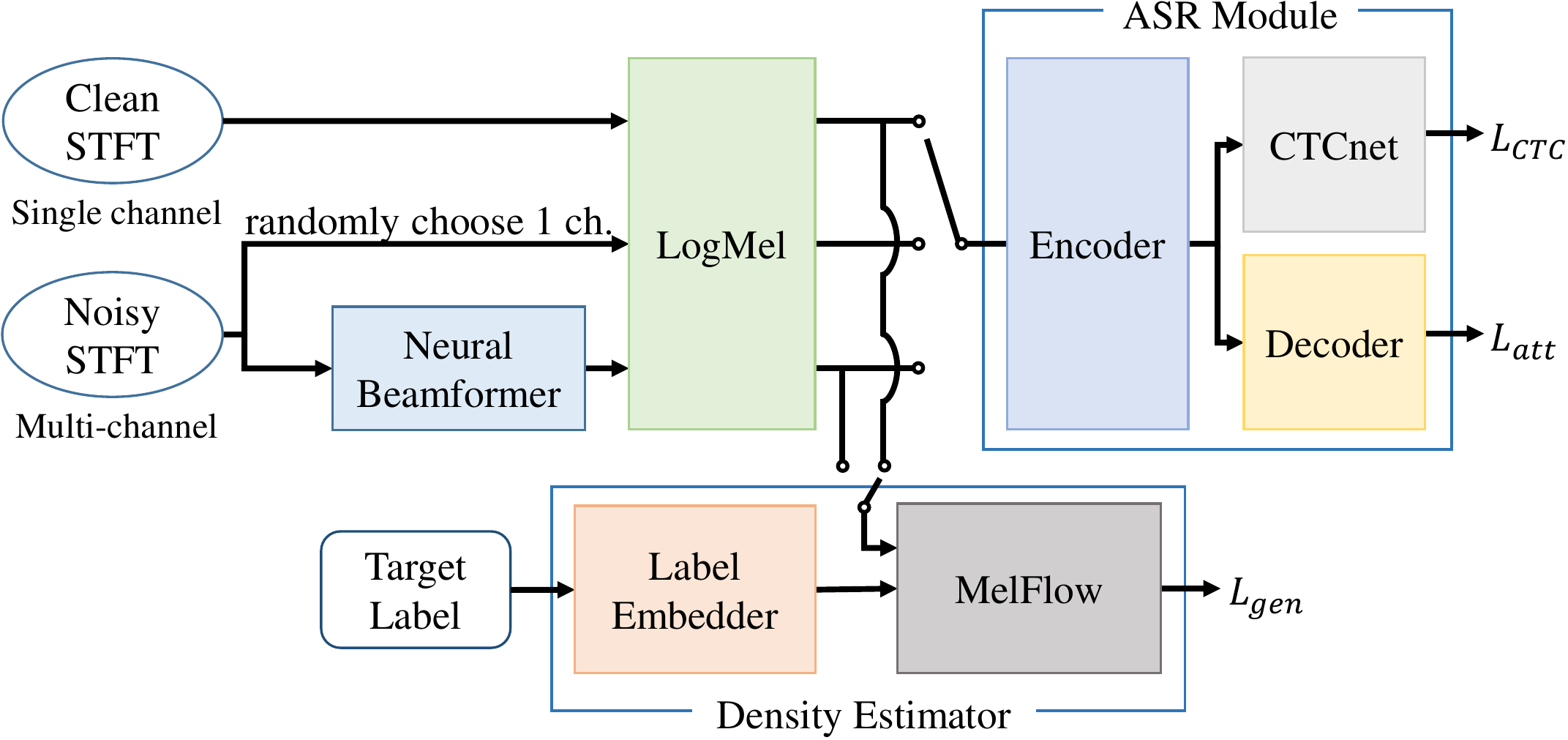}
	\label{fig:fig1a}
	}
	\centering
	\subfigure[]{
	\includegraphics[height=0.35\linewidth]{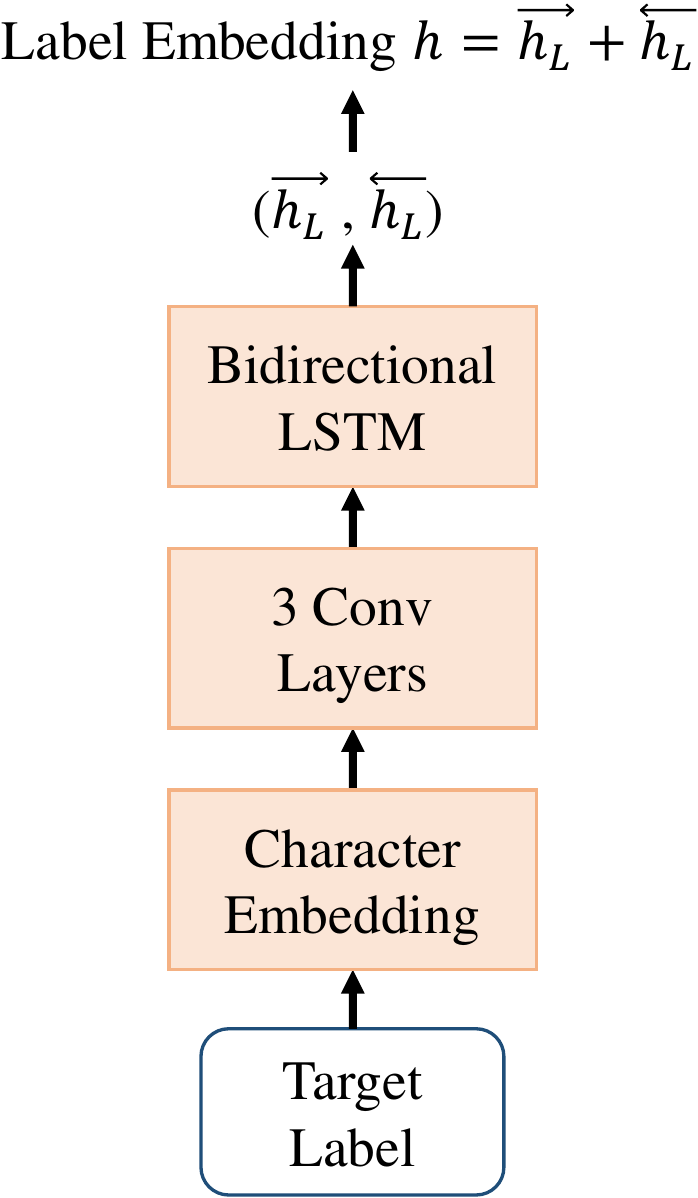}
	\label{fig:fig1b}
	}
    \caption{(a) The overall structure of the proposed model. (b) The structure of Label Embedder.  
    }
	\label{fig:proposed_model}
\end{figure*}

Here, we propose to incorporate a flow-based density estimation task within the multi-channel end-to-end ASR. Our insight is that the ASR objective is insufficient for fully training the front-end since it only focuses on preserving the phonetic information. For the robust front-end,  we now exploit the distribution of the target clean speech on which the original ASR system is trained. 

We also present a novel flow-based generative model \textit{MelFlow} for estimating the likelihood of mel-spectrograms. Many flow-based models for estimating the distribution of raw audio have been studied~\cite{prenger2019waveglow,ping2018clarinet}, but have not been applied to mel-specograms. We introduce a new flow-based model for mel-spectrograms and explain the architecture of MelFlow.

\subsection{Flow-based Generative Model}
A flow-based model is a generative model which consists of a stack of invertible mappings from a simple distribution $p_{Z}(\boldsymbol{z})$ to a complex distribution $p_{X}(\boldsymbol{x})$~\cite{dinh2016density}. Let $\boldsymbol{f}_{i}$ be a mapping from $\boldsymbol{z}^{i-1}$ to $\boldsymbol{z}^{i}$ , $\boldsymbol{z}^{0} = \boldsymbol{x}$ and $\boldsymbol{z}^{n} = \boldsymbol{z}$ ($\boldsymbol{z}^{i} \in \mathbb{R}^D$ for $i=0, ... , n$). Then $\boldsymbol{x} \in \mathbb{R}^D$ is transformed into $\boldsymbol{z}$ through a chain of invertible mappings:
\begin{equation}
\label{eq:eq8}
\boldsymbol{z} = \boldsymbol{f}_{n} \circ \boldsymbol{f}_{n-1} \circ ... \circ \boldsymbol{f}_{1}(\boldsymbol{x}).
\end{equation}
By change of variables theorem, the log-likelihood of data $\boldsymbol{x}$ is expressed as follows:
\begin{equation}
\label{eq:eq9}
\log p_{X}(\boldsymbol{x}) = \log p_{Z}(\boldsymbol{z}) + \sum_{i=1}^{n}\log \left|\det\left(\frac{\partial \boldsymbol{f}_{i}}{\partial\boldsymbol{z}^{i-1}}\right)\right|.
\end{equation}
By maximizing $\log p_{X}(\boldsymbol{x})$, we obtain a density estimator of data $\boldsymbol{x}$. A typical choice for the prior distribution $p_{Z}(\boldsymbol{z})$ is a zero-mean isotropic multivariate Gaussian $\mathcal{N}(\boldsymbol{0}, \boldsymbol{I})$. If $\boldsymbol{z}$ is obtained by Eq.~\eqref{eq:eq8}, the first term $\log p_{Z}(\boldsymbol{z})$ in Eq.~\eqref{eq:eq9} can be calculated easily. However, it is too expensive to compute the second term $\sum_{i=1}^{n}\log \left|\det\left(\frac{\partial \boldsymbol{f}_{i}}{\partial\boldsymbol{z}^{i-1}}\right)\right|$ directly. To reduce the computational complexity, $\boldsymbol{f}_{i}$ is required to have a tractable Jacobian. The affine coupling layer~\cite{dinh2016density} satisfies such a requirement and is defined as follows:
\begin{equation}
\label{eq:eq10}
\boldsymbol{z}^{i}_{a} = \boldsymbol{z}^{i-1}_{a},
\end{equation}
\begin{equation}
\label{eq:eq11}
\boldsymbol{z}^{i}_{b} = \boldsymbol{z}^{i-1}_{b} \odot \exp{(\boldsymbol{\sigma}(\boldsymbol{z}^{i-1}_{a}))} + \boldsymbol{\mu}(\boldsymbol{z}^{i-1}_{a}),
\end{equation}
where $\boldsymbol{z}^{i}_{a} \in \mathbb{R}^{\frac{D}{2}}$ is the first half, $\boldsymbol{z}^{i}_{b} \in \mathbb{R}^{\frac{D}{2}}$ is the last half of $\boldsymbol{z}^i$, $\boldsymbol{\mu}$() and $\boldsymbol{\sigma}$() are the functions from $\mathbb{R}^{\frac{D}{2}} \mapsto \mathbb{R}^{\frac{D}{2}}$, and $\odot$ stands for the element-wise product. The Jacobian matrix of the affine coupling layer is a lower triangular matrix and $\log \left|\det\left(\frac{\partial \boldsymbol{f}_{i}}{\partial\boldsymbol{z}^{i-1}}\right)\right|$ can be computed efficiently:
\begin{equation}
\label{eq:eq12}
\log \left|\det\left(\frac{\partial\boldsymbol{f}_{i}}{\partial\boldsymbol{z}^{i-1}}\right)\right| = \sum^\frac{D}{2}_{j=1}\boldsymbol{\sigma}(\boldsymbol{z}^{i-1}_{a})_j,
\end{equation}
where $\boldsymbol{\sigma}(\boldsymbol{z}^{i-1}_{a})_j$ is the $j$-th elemtent of $\boldsymbol{\sigma}(\boldsymbol{z}^{i-1}_{a})$.

\subsection{MelFlow}
\begin{figure}[t]
	\centering
	\includegraphics[width=7.5cm]{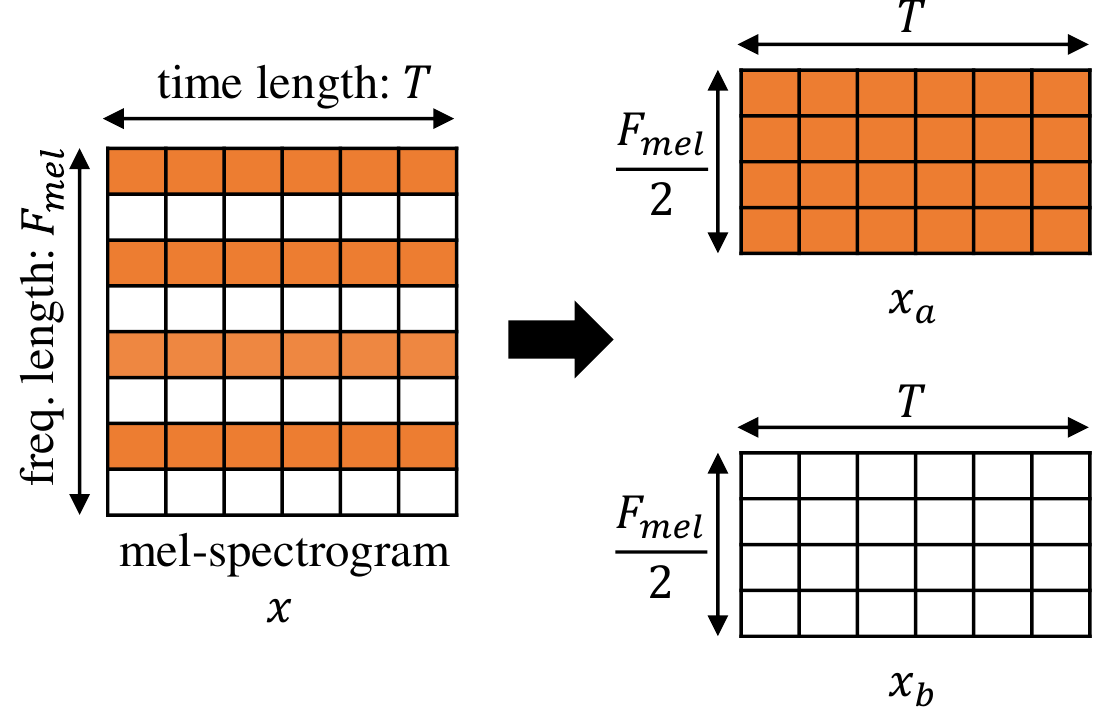}
	\caption{Splitting a mel-spectrogram into $\boldsymbol{x}_{a}$ and $\boldsymbol{x}_{b}$} 
	\label{fig:splitting}
\end{figure}
\begin{figure}[t]
	\centering
	\includegraphics[width=7.5cm]{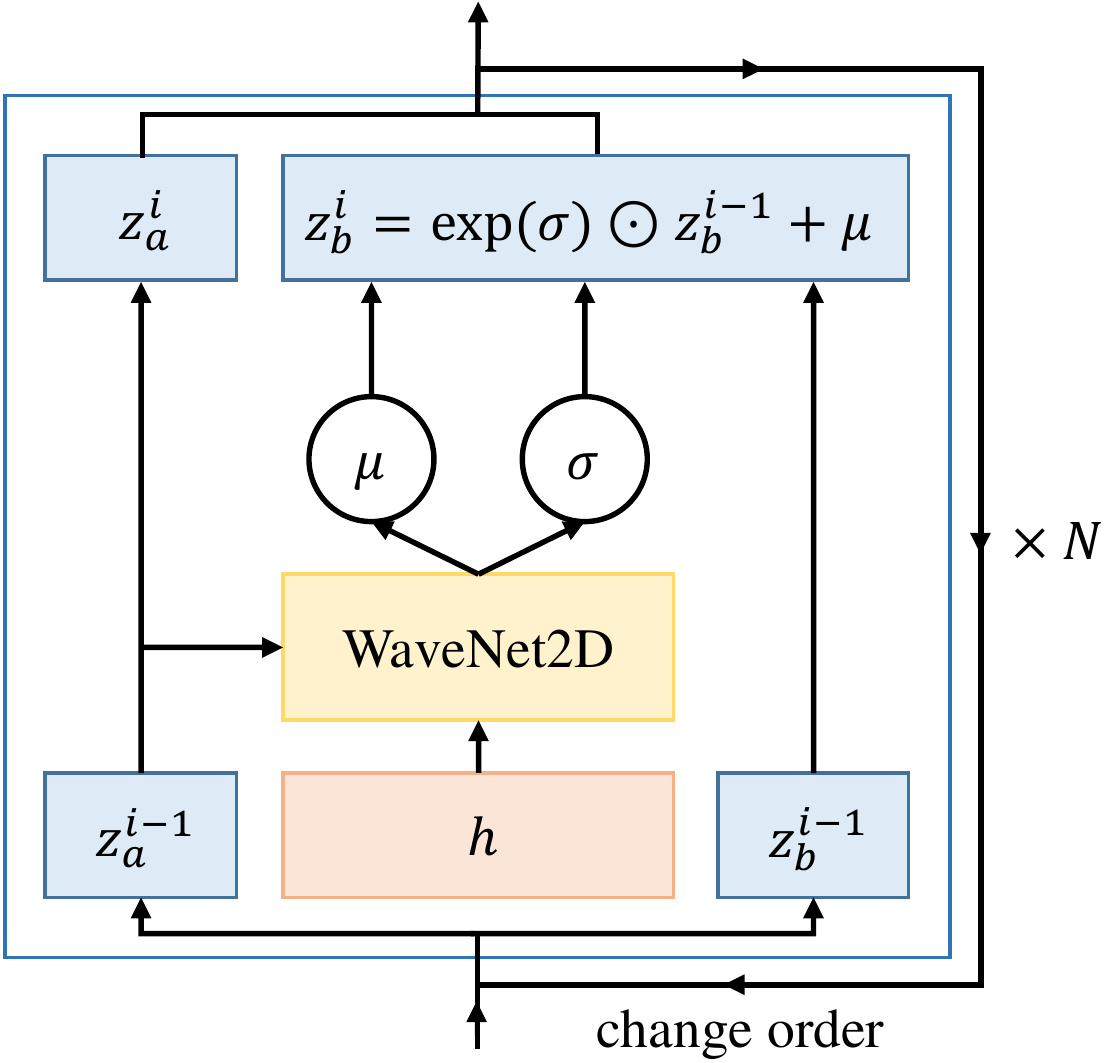}
	\caption{MelFlow network} 
	\label{fig:MelFlow}
\end{figure}
We now turn to building a density estimator for mel-spectrograms. An input data $\boldsymbol{x} \in \mathbb{R}^{F_{mel} \times T}$ is a mel-spectrogram where $F_{mel}$ is a fixed frequency bin length of the mel-spectrograms and $T$ is a variable time-length depending on an utterance. It is splited into $\boldsymbol{x}_{a} \in \mathbb{R}^{\frac{F_{mel}}{2} \times T}$ and $\boldsymbol{x}_{b} \in \mathbb{R}^{\frac{F_{mel}}{2} \times T}$ along the frequency axis, which is depicted in Figure~\ref{fig:splitting}. Let $\boldsymbol{z}^{0}_{a} = \boldsymbol{x}_{a}$ and $\boldsymbol{z}^{0}_{b} = \boldsymbol{x}_{b}$. Throughout one of flow stacks, $\boldsymbol{z}^{i-1}_{a}$ and $\boldsymbol{z}^{i-1}_{b}$ are transformed in a different way: $\boldsymbol{z}^{i-1}_{a}$ remains and $\boldsymbol{z}^{i-1}_{b}$ is transformed into $\boldsymbol{z}^{i}_{b}$ as follows
\begin{equation}
\label{eq:eq13}
(\boldsymbol{\mu}, \boldsymbol{\sigma}) = \WaveNet(\boldsymbol{z}^{i-1}_{a}),
\end{equation}
\begin{equation}
\label{eq:eq14}
\boldsymbol{z}^{i}_{a} = \boldsymbol{z}^{i-1}_{a},
\end{equation}
\begin{equation}
\label{eq:eq15}
\boldsymbol{z}^{i}_{b} = \exp{(\boldsymbol{\sigma})} \odot \boldsymbol{z}^{i-1}_{b} + \boldsymbol{\mu},
\end{equation}
where WaveNet2D can be any function $\mathbb{R}^{\frac{F_{mel}}{2} \times T} \mapsto \mathbb{R}^{F_{mel} \times T}$ and a multiplicative term $\boldsymbol{\sigma} \in \mathbb{R}^{\frac{F_{mel}}{2} \times T}$ and an additive term $\boldsymbol{\mu} \in \mathbb{R}^{\frac{F_{mel}}{2} \times T}$ depend only on $\boldsymbol{x}_{a}$. In this work, we use multiple layers of dilated 2D-convolutions with gated-tanh nonlinearities, residual connections and skip connections for WaveNet2D. WaveNet2D is similar to WaveNet~\cite{oord2016wavenet}, but different in that WaveNet2D is composed of non-causal 2D-convolutions. The Jacobian determinant in Eq.~\eqref{eq:eq12} is computed as follows:
\begin{equation}
\label{eq:eq16}
\log \left|\det\left(\frac{\partial\boldsymbol{f}_{i}}{\partial\boldsymbol{z}_{i-1}}\right)\right| = \sum^{\frac{F_{mel}}{2}}_{f=1}\sum^{T}_{t=1}\boldsymbol{\sigma}_{f,t}.
\end{equation}
MelFlow achieves a more flexible and high-expressive model by stacking multiple flow operations as illustrated in Figure~\ref{fig:MelFlow}. Also, we change the order of $\boldsymbol{z}^{i}_{a}$ and $\boldsymbol{z}^{i}_{b}$ before each flow operation (i.e. $\boldsymbol{z}^{i}_{a}$ and $\boldsymbol{z}^{i}_{b}$ are transformed by Eq.~\eqref{eq:eq15} alternately). 

We can further improve the model by exploiting target labels. Note that we use the density estimator to train the front-end and target labels are available during training. As shown in Figure~\ref{fig:fig1b}, a target label is first embedded into a sequence of vectors through Character Embedding. A stack of 3 convolutional layers is applied to the sequence of vectors and the output is passed into a bidirectional LSTM. A label embedding $\boldsymbol{h}$ is finally obtained by summing the last hidden state of the forward path $\overrightarrow{\boldsymbol{h}_{L}}$ and the backward path $\overleftarrow{\boldsymbol{h}_{L}}$ in the bidirectional LSTM:
\begin{equation}
\label{eq:eq17}
\boldsymbol{h} = \overrightarrow{\boldsymbol{h}_{L}} + \overleftarrow{\boldsymbol{h}_{L}}.
\end{equation}
Label Embedder can be considered as a compact version of the encoder in~\cite{shen2018natural}. The attention mechanism isn't included due to the restriction of GPU memory which is mostly occupied by the ASR module. We now reformulate Eq.~\eqref{eq:eq13} by adding a global condition $\boldsymbol{h}$ to WaveNet2D as follows:
\begin{equation}
\label{eq:eq18}
(\boldsymbol{\sigma}, \boldsymbol{\mu}) = \WaveNet(\boldsymbol{z}^{i-1}_{a}, \boldsymbol{h}).
\end{equation}

Considering the fact that $T$ is a variable time-length, the generative loss $L_{gen}$ (the negative log-likelihood, NLL) is defined as:
\begin{equation}
\label{eq:eq19}
L_{gen} = -\frac{1}{TF_{mel}}\log p_{X}(\boldsymbol{x}),
\end{equation}
where $F_{mel}$ is a fixed frequency-bin length and $T$ varies by the input data $\boldsymbol{x}$.

\subsection{Joint Training with Density Estimation}
The proposed model incorporates the density estimator to the base ASR architecture as shown in Figure~\ref{fig:fig1a}.
\begin{algorithm}[tb]
\caption{A joint training step with density estimation}
\label{alg:algorithm}
\textbf{Input}: A mini-batch $S$, a target label $L$, the neural beamformer $M_{NB}$, the density estimator $M_{DE}$ and the ASR module $M_{ASR}$

\begin{algorithmic}
\IF{$S$ consists of $clean\;speech$ (single channel)}
\STATE $L_{ASR} \leftarrow$ compute the ASR loss with $(S, L, M_{ASR})$\\
\STATE $L_{gen} \leftarrow$ compute the NLL with $(S, L, M_{DE})$\\
\STATE optimize $M_{ASR}$ and $M_{DE}$ with $(L_{ASR}, L_{gen})$
\ELSIF{$S$ consists of $noisy\;speech$ (multi-channel)}
\STATE $S_{enh} \leftarrow$ enhance by passing $S$ through $M_{NB}$\\
\STATE sample $u \sim unif(0,1)$
\IF{$u < 0.5$}
\STATE $S_{rand} \leftarrow$ choose 1 channel randomly from $S$
\STATE $(S_{ASR}, S_{gen}) \leftarrow (S_{rand}, S_{enh})$
\ELSE
\STATE $(S_{ASR}, S_{gen}) \leftarrow (S_{enh}, S_{enh})$
\ENDIF
\STATE $L_{ASR} \leftarrow$ compute the ASR loss with $(S_{ASR}, L, M_{ASR})$\\
\STATE $L_{gen} \leftarrow$ compute the NLL with $(S_{gen}, L, M_{DE})$\\
\STATE optimize $M_{ASR}$ and $M_{NB}$ with $(L_{ASR}, L_{gen})$
\ENDIF
\end{algorithmic}
\end{algorithm}
We take advantage of non-parallel clean and noisy speech data by employing the density estimation task. To be specific, when clean speech data comes in a mini-batch, the ASR module and the density estimator are trained ordinarily. If a noisy speech data comes in the mini-batch, cases are divided into two. In the first case with a probability $0.5$, the ASR module receives a randomly chosen channel of the noisy speech. In the other case, the ASR module receives enhanced speech from the neural beamformer. The ASR module and the neural beamformer are trained for both cases while the density estimator is only used for computing $L_{gen}$. Algorithm~\ref{alg:algorithm} describes a joint training step with density estimation.

\section{Experiments}
In order to evaluate the proposed method in a noisy speech scenario, we conducted a set of experiments using the CHiME-4 dataset.
\subsection{Dataset}
CHiME-4 is a speech recognition dataset which is recorded by a multi-microphone tablet device in every day, noisy environments. The tablet device is equipped with 6-channel microphones where 5 of them face forward and the other one faces backward. In this work, we excluded the speech data recorded by the microphone facing backward; hence the number of channels $C$ was 5. CHiME-4 employs two types of data: (i) speech data recorded in real noisy environments (i.e., on a bus, cafe, pedestrian area, and street junction), and (ii) simulated speech data that is generated by manually mixing clean speech data with background noise. Also, the dataset is divided into training, development and evaluation sets. The training set consists of 3 hours of real noisy utterances from 4 speakers and 15 hours of simulated noisy utterances from 83 speakers. The development set consists of 2.9 hours of real and simulated noisy utterances from 4 speakers, respectively. The evaluation set consists of 2.2 hours of utterances for each real and simulated noisy data. 

We also employed Wall Street Journal (WSJ) read speech for single channel clean speech dataset. WSJ's si-284 set contains 82 hours of clean utterances and was used only for training the model.

\subsection{Model Configurations}

\paragraph{Neural Beamformer.}
To compute 200 STFT coefficients (i.e., $F$=201), the 25ms-width Hanning window with a 10ms shift was used. We used a 3-layer bidirectional LSTM with 300 cells for BiLSTM in Eq.~\eqref{eq:eq3}. Also, a linear projection layer with 300 units was inserted after every layer of bidirectional LSTM. For FClayer in Eq.~\eqref{eq:eq4}, a 1-layer linear transformation was used. To estimate the reference microphone, a 2-layer linear transformation was used with tanh activation. The reference microphone vector $\boldsymbol{r}$ was finally estimated using the softmax function.

\paragraph{LogMel.}
STFT coefficients were converted to mel-spectrograms by LogMel. The mel-scale is primarily used to mimic the non-linear human ear perception of sound. In our experiments, $F_{mel}$ was 80.

\paragraph{Label Embedder.}
We used a 16-dimensional character embedding. The kernel sizes of 1D convolutional layers were set to be 3 and the sizes of input and output were the same as 16. The ReLU activation and the batch normalization were used at the end of each convolutional layer. We stacked 3 convolutional layers. The sizes of the hidden state in the bidirectional LSTM were 256 and the 2 last hidden states of the forward and backward paths were summed to obtain the label embedding $h \in \mathbb{R}^{h_{dim}}$ where $h_{dim}$ was set to be 256.

\paragraph{MelFlow.}
We used MelFlow consisting of 8 affine coupling layers. For each WaveNet2D, the kernel sizes for the first and last convolutional layer were set to be 1. The rest of the layers (i.e., middle 4 layers) was composed of 20 channels and kernel with size 3, and used for residual connections, skip connections and gated-tanh unit. For conditioning the label embedding $h$ globally, a fully connected layer was included in WaveNet2D. All the weights of the last convolutional layers in WaveNet2D were initialized to be zero. This initialization has been known to result in the stable training procedure. 
\paragraph{ASR module.}
For Encoder, a 4-layer 2D convolutional network and a 3-layer bidirectional LSTM with 1024 cells were used. The kernel sizes were set to be (3,3) for all layers in the convolutional network and channels were set to be (1, 64), (64, 64), (64, 128) and (128, 128), respectively. A linear projection layer with 1024 units was inserted after every layer of bidirectional LSTM in Encoder. To boost the ASR optimization, we adopted a joint CTC-attention loss function. For CTCnet, we used a 1-layer linear transformation with output dimension 52 indicating characters. For Decoder, a unidirectional LSTM with 1024 cells and a 1-layer linear transformation were used. To connect Encoder and Decoder, we leveraged the attention mechanism.
\paragraph{ASR loss.}
When the ASR module is trained with only the attention loss, it usually suffers from misalignment because the attention mechanism is too flexible to predict the right alignments. It has been reported that the CTC loss enforces monotonic alignments between speech and label sequences due to the left-to-right constraint~\cite{kim2017joint}. Thus the auxiliary CTC loss helps the attention model to have proper alignments and boosts the whole training procedure. The CTC loss $L_{CTC}$ can be calculated efficiently with the forward-backward algorithm and the attention loss $L_{att}$ is also easily obtained with a teacher forcing method at Decoder. The joint CTC-attention objective $L_{ASR}$ is expressed as follows with a tuning parameter $\lambda$:
\begin{equation}
\label{eq:eq20}
L_{ASR} = \lambda  L_{CTC} + (1-\lambda) L_{att},
\end{equation}
where we set $\lambda$ to 0.5 for the experiments.
\paragraph{Total loss.}
The total loss $L_{tot}$ is defined as:
\begin{equation}
\label{eq:eq21}
L_{tot} = L_{ASR} + \beta L_{gen},
\end{equation}
where $\beta$ is a hyperparameter. We experimented with different values of $\beta$.

\paragraph{Baseline.}
We used ESPnet as the baseline. The baseline doesn't have Label Embedder and MelFlow in Figure~\ref{fig:fig1a}. All the other configurations were the same as the proposed model.

\section{Results}

\begin{table*}[t]
\centering
\begin{tabular}{@{}lclccccl@{}}
\toprule
 & \multicolumn{1}{l}{}                  & \multicolumn{1}{c}{}        & \multicolumn{2}{c}{development set} & \multicolumn{2}{c}{evaluation set} &  \\ \midrule
 & \multicolumn{1}{l}{}                  & \multicolumn{1}{c}{Model}   & simulated data    & real data       & simulated data   & real data       &  \\ \midrule
 & \multicolumn{1}{l}{}                  & Baseline                    & 9.1               & 9.2             & 13.6             & 17.2            &  \\ \midrule
 & \multirow{4}{*}{w/o label condition}  & Proposed Model ($\beta$ = 1)    & 8.9               & 9.5             & 13.2             & 17.3            &  \\
 &                                       & Proposed Model ($\beta$ = 0.25) & 8.8               & 9.1             & \textbf{12.7}    & 17.0            &  \\
 &                                       & Proposed Model ($\beta$ = 0.1)  & 8.7               & 9.1             & 13.2             & 17.4            &  \\
 &                                       & Proposed Model ($\beta$ = 0.01) & 9.1               & \textbf{8.9}    & 13.2             & 17.2            &  \\ \midrule
 & \multirow{4}{*}{with label condition} & Proposed Model ($\beta$ = 1)    & 8.6               & 9.1             & 12.9             & 16.7            &  \\
 &                                       & Proposed Model ($\beta$ = 0.25) & \textbf{8.1}      & 9.0             & 13.1             & 16.8            &  \\
 &                                       & Proposed Model ($\beta$ = 0.1)  & 8.5               & 9.1             & 13.2             & 16.7            &  \\
 &                                       & Proposed Model ($\beta$ = 0.01) & 8.4               & \textbf{8.9}    & 13.3             & \textbf{16.3}   &  \\ \bottomrule
\end{tabular}
\caption{Word error rate [\%] on CHiME-4 dataset}
\label{tab:wer}
\end{table*}

\begin{table}[t]
\centering
\begin{tabular}{@{}llcccl@{}}
\toprule
 &                             & SDR   & ESTOI & PESQ &  \\ \midrule
 & Baseline                    & 15.75 & 0.83  & 1.87 &  \\ \midrule
 & Proposed Model ($\beta$ = 1)    & 14.44 & 0.82  & 1.83 &  \\
 & Proposed Model ($\beta$ = 0.25) & 15.78 & 0.83  & 1.87 &  \\
 & Proposed Model ($\beta$ = 0.1)  & 15.85 & 0.83  & 1.88 &  \\
 & Proposed Model ($\beta$ = 0.01) & 15.87 & 0.83  & 1.88 &  \\ \bottomrule
\end{tabular}
\caption{Speech enhancement scores on CHiME-4 simulated evaluation set}
\label{tab:enhancement}
\end{table}

We compared the noisy speech recognition performances of the baseline and the proposed model on the CHiME-4 dataset. The baseline was trained with only the ASR objective $L_{ASR}$. We used 2 types of the proposed model in the experiment: one with both Label Embedder and MelFlow and the other with MelFlow. Also, various values of the hyperparameter $\beta$ in Eq.~\eqref{eq:eq21} were used in the experiments: 1, 0.25, 0.1 and 0.01. Attention scores and CTC scores were averaged at a ratio of 7:3 and a beam search algorithm with the beam size 20 was used for decoding. An RNN-based language model was also used to enhance the quality of speech recognition. Word error rates (WERs) of the outputs of the different models are shown in Table \ref{tab:wer}. Overall, the proposed model without the label condition showed better performances than the baseline. For $\beta = 0.25$, the proposed model outperformed the baseline with an absolute decrease of 0.9\% in terms of WER on the simulated noisy data in the evaluation set. However, the improvement was not obvious over the real noisy data in both development and evaluation sets. When the label condition was incorporated into the model, the overall performance showed significant improvement and, surprisingly, the WERs of the proposed model were improved in all cases. For $\beta = 0.01$, the average WER on the real noisy data in the evaluation set achieved 16.3\%. The experiment demonstrates that the auxiliary objective from the density estimation task leads the front-end to learn more general representations and this leads to the improved performance of noisy speech recognition. The difference of performances between the models with/without the label condition suggests that the accurate density estimator should be used in order that the front-end gets more benefits from the generative loss.

One may ask whether the proposed model achieves improvements in respect of speech enhancement. Unfortunately, the answer is no. Speech enhancement scores are illustrated in Table \ref{tab:enhancement}. We evaluated speech-to-distortion ratio (SDR~\cite{vincent2006performance}), extended short-time objective intelligibility (ESTOI~\cite{jensen2016algorithm}), and perceptual evaluation of speech quality (PESQ~\cite{rix2001perceptual}) between the enhanced speech and the reference speech in the evaluation set. The CHiME-4 dataset provides the clean data recorded by the close-talk microphone and we used this data as the reference speech. We used the proposed model with the label condition for the speech enhancement evaluation. The overall scores of the proposed model were almost same as the ones of the baseline. This implies that in the proposed model the representation after the front-end is generalized and useful for the ASR module but this doesn't necessarily mean the improvement of the metrics of speech enhancement. Multi-task learning of speech enhancement and density estimation could be beneficial for the speech enhancement scores and we leave it for future work.

\section{Conclusion}
In this work, we presented the novel method which employs flow-based density estimation for robust multi-channel ASR. We also proposed \textit{MelFlow} to estimate the distribution of mel-spectrograms of clean speech. In the experiments, we demonstrated that the proposed model shows better performance than the conventional ASR model in terms of word error rate (WER) on noisy multi-channel speech data. We verified that the auxiliary generative objective helps the front-end to learn more regularized representations which lead to improvements on noisy speech recognition. 

For future work, we will apply an autoregressive model or a Gaussian mixture model (GMM) to estimate the probability density on behalf of MelFlow. Also, we will apply our joint training scheme with density estimation to speech enhancement.

\section*{Acknowledgments}

This work was supported by Samsung Research Funding Center of Samsung Electronics under Project Number SRFC-IT1701-04.

\bibliography{reference}
\bibliographystyle{named}

\end{document}